\documentclass[english, twocolumn, 10pt, aps, superscriptaddress, floatfix, showpacs, prl, citeautoscript]{revtex4-1}

\usepackage{graphicx}
\usepackage{dcolumn}
\usepackage{bm}
\usepackage{amssymb}

\begin{document}

\title{Second-harmonic current-phase relation in Josephson junctions with ferromagnetic barriers}

\author{M.J.A. Stoutimore}
 \affiliation{Department of Physics, University of Illinois at Urbana-Champaign, Urbana, IL 61801, USA}

\author{A.N. Rossolenko}
\affiliation{Institute of Solid State Physics, Russian Academy of Sciences, Chernogolovka, 142432, Russia}

\author{V.V. Bolginov}
\affiliation{Institute of Solid State Physics, Russian Academy of Sciences, Chernogolovka, 142432, Russia}
\affiliation{Skobeltsyn Institute of Nuclear Physics, Lomonosov Moscow State University, Moscow 119991, Russia}
\affiliation{ Russian National University of Science and Technology (NUST) MISiS, 4 Leninsky Prospect, Moscow 119049, Russia }

\author{V.A. Oboznov}
\affiliation{Institute of Solid State Physics, Russian Academy of Sciences, Chernogolovka, 142432, Russia}

\author{A.Y. Rusanov}
\affiliation{Institute of Solid State Physics, Russian Academy of Sciences, Chernogolovka, 142432, Russia}

\author{D.S. Baranov}
\affiliation{Institute of Solid State Physics, Russian Academy of Sciences, Chernogolovka, 142432, Russia}
\affiliation{Moscow Institute of Physics and Technology, Dolgoprudny, 141700, Russia}

\author{N. Pugach}
\affiliation{Skobeltsyn Institute of Nuclear Physics, Lomonosov Moscow State University, Moscow 119991, Russia}
\affiliation{MIEM, National Research University Higher School of Economics, 101000 Moscow, Russia}

\author{S.M. Frolov}
\affiliation{Department of Physics and Astronomy, University of Pittsburgh, Pittsburgh, PA 15260, USA}

\author{V.V. Ryazanov}
\affiliation{Institute of Solid State Physics, Russian Academy of Sciences, Chernogolovka, 142432, Russia}
\affiliation{ Russian National University of Science and Technology (NUST) MISiS, 4 Leninsky Prospect, Moscow 119049, Russia }
\affiliation{Faculty of Physics, National Research University Higher School of Economics, 101000 Moscow, Russia}

\author{D.J. Van Harlingen}
 \affiliation{Department of Physics, University of Illinois at Urbana-Champaign, Urbana, IL 61801, USA}

\date{\today}

\begin{abstract}
We report the observation of a current-phase relation dominated by the second Josephson harmonic in superconductor-ferromagnet-superconductor junctions. The exotic current-phase relation is realized in the vicinity of a temperature-controlled 0-to-$\pi$ junction transition, at which the first Josephson harmonic vanishes. Direct current-phase relation measurements, as well as Josephson interferometry, non-vanishing supercurrent and half-integer Shapiro steps at the 0-$\pi$ transition self-consistently point to an intrinsic second harmonic term, making it possible to rule out common alternative origins of half-periodic behavior. While surprising for diffusive multimode junctions, the large second harmonic is in agreement with theory predictions for thin ferromagnetic interlayers.
\end{abstract}

\maketitle

The sinusoidal dependence of supercurrent on the phase difference across the junction $\phi$ was originally derived for superconductor-insulator-superconductor junctions, but has for a long time been used to describe most of the experimentally realized junctions \cite{Josephson}. Advances in materials science and nanofabrication have lead to the observation of a large variety of current-phase relations (CPRs) \cite{Golubov-review}. For example, $\pi$-junctions may still have a sinusoidal CPR but with a phase shift of $\pi$ \cite{Ryazanov-junction,Klapwijk-junction,cleuziounatnano06,vandamnature06}. 
$\phi_0$-junctions violate time-reversal symmetry with a phase shift $\phi_0$ other than 0 or $\pi$ in the CPR, meaning that their current-phase relations have no phase-inversion symmetry \cite{GoldobinPRB07, szombati2015josephson}. 
Narrow and/or ballistic weak links with non-sinusoidal current-phase relations, i.e. containing higher sine components, have been reported based on a variety of materials \cite{della2007measurement,sochnikovPRL15,nanda2017current,EnglishPRB16,spanton2017current}.  Finally, fractional current-phase relations such as $\mathrm{sin(\phi/2)}$ are being searched for in topological superconductor junctions \cite{kitaev2001unpaired, wiedenmann20164pi, deacon2017josephson}. These developments motivate new studies of exotic CPR.
    
This paper is focused on a junction with a rare second harmonic current-phase relation dominated by the $\mathrm{sin(2\phi)}$ contribution.  In contrast with previous studies of higher Josephson components, the junction barrier is a diffusive metal with a macroscopic number of modes. The second harmonic CPR is realized in the vicinity of a temperature-controlled 0-$\pi$ transition of a superconductor-ferromagnet-superconductor (SFS) junction. At the transition temperature T$_\pi$ the amplitude of the first order $\mathrm{sin(\phi)}$ term goes through zero in order to change sign. If a significant higher-order term is present it can become the leading 
one \cite{Buzdin-review}. Earlier studies of the  0-$\pi$ transition in various junctions have suggested a second-order CPR \cite{Klapwijk-squid, Sellier-Sin2phi, cleuziounatnano06},
however those experiments could not rule out alternative explanations for $\pi$-periodic behavior such as due to more than one junction in the loop, disorder in the junction or driven phase dynamics \cite{Frolov-Shapiro}.

\begin{figure}[ht]
  \includegraphics[width=\columnwidth]{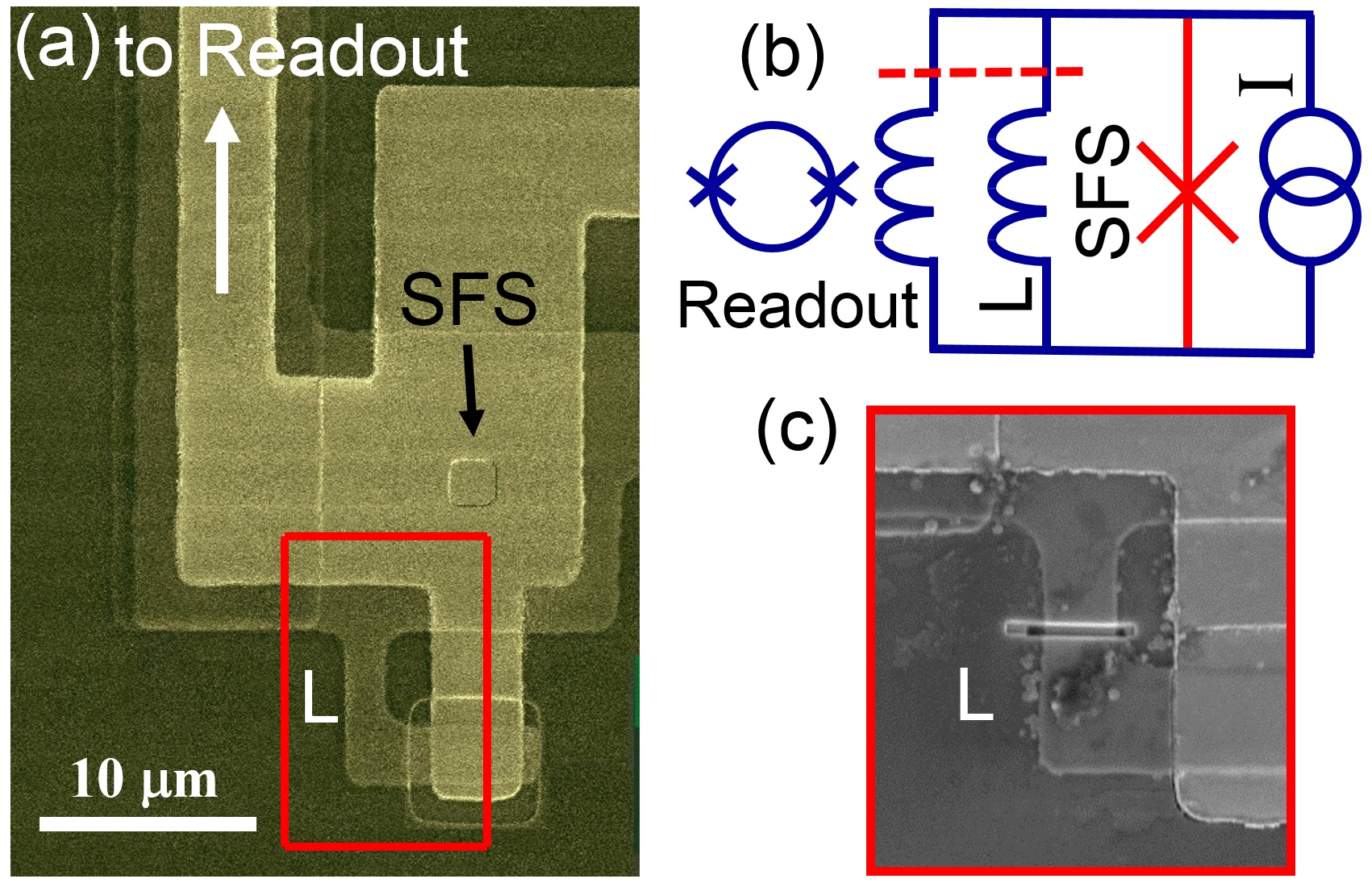}
  \caption{
(a) Optical micrograph of a prototype device, zoomed in on the area containing the trilayer SFS junction, as well as the small shunt inductor $L$. Red frame corresponds to the area in panel (c).
(b) schematic of the CPR measurement. Dashed line indicates that both inductors are later cut to perform current-voltage measurements on the SFS junction. (c) scanning electron micrograph of the measured device, focused on the region marked by red in panel (a) after the inductor $L$ is cut with a focused ion beam.
 }
 \label{fig1}
\end{figure}
    
Here we perform four distinct measurements on a single SFS junction, all four indicating a dominant and intrinsic second-order Josephson effect. First, a direct measurement of the current-phase relation is performed by embedding a single SFS junction into a superconducting loop. In this measurement a second harmonic manifests as doubling of the superconducting loop response modulation near the 0-$\pi$ transition. Subsequently, the loop is cut and Josephson interferometry is performed on the same junction showing Fraunhofer-like patterns with half-flux quantum periodicity near $T_\pi$. Third, the same junction is found to exhibit a non-vanishing critical current at the 0-$\pi$ transition. And fourth, half-integer Shapiro steps are observed around the 0-$\pi$ transition. All effects are consistent with a positive sin(2$\phi$) term with the critical current density of $\approx400-600 A/cm^2$. We find this to be in agreement with theory developed for diffusive junctions \cite{BuzdinPRB05}.
  
For the junction barrier we use a Cu$_{47}$Ni$_{53}$ alloy (in atomic percentage) which is a weak ferromagnet with a Curie temperature of approximately 60 K and a rigid out-of-plane domain structure \cite{veshchunovJETPLett07}. SFS junctions were fabricated by depositing a Nb-CuNi-Nb trilayer in a single vacuum cycle using argon sputtering followed by multistep fabrication process described in Ref.[\onlinecite{bolginovJLTP18}]. The junction studied in the main text has a barrier thickness of $d_F$~=~7.3~nm, and an area of $(2 \times 2\pm0.5)\mu m^2$ (Fig. 1(a)). Relative to Refs. \cite{Frolov-Shapiro,Oboznov-PRL,Frolov-CPR}, the trilayer fabrication process resulted in a lower barrier thickness of the first 0-$\pi$ transition \cite{bolginovJLTP18}, which led to the increased second harmonic amplitude.

For the direct CPR measurement, the SFS junction is shorted by a parallel combination of two superconducting Nb loops, the millimeter-scale readout loop with an inductance $L_{readout}$ and a micron-scale loop with an inductance $L$ (see 
Fig. 1(b) and supplementary materials). The effective inductance of the device is close to $L$. The readout inductor is coupled to a commercial dc SQUID sensor which detects flux $\Phi$ in the readout loop. The bias current $I$ is applied across the SFS junction, and inductors $L$ and $L_{readout}$ in parallel, it divides between the three branches to satisfy fluxoid quantization. CPR information is extracted from $\Phi(I)$. After performing CPR measurements, both $L$ and $L_{readout}$ are cut for the current-voltage measurements on the same junction (Fig. 1(c)).

The value of L$_{readout} \approx 1.2 nH$  is chosen in order to optimally couple the sample to a commercial readout SQUID. If $L_{readout}$ were the only inductor in the circuit the device would always be in the strongly hysteretic regime with multi-valued $\Phi(I)$ that makes it impossible to extract CPR  \cite{Barone}. This is due to the high critical current density in SFS junctions with thin barriers. The small inductor $L$ is designed to suppress the hysteresis of $\Phi(I)$. The single valued $\Phi(I)$ dependence is expected for junctions with purely first-order CPR when the parameter $\beta_{L1}= 2\pi I_{c1}L/\Phi_0<1$, where $\Phi_0$ is the superconducting flux quantum, $I_{c1}$ is the supercurrent amplitude of the first Josephson harmonic. For a Josephson junction with a purely second harmonic CPR, the condition is more stringent: $\beta_{L2}= 2\pi I_{c2}L/\Phi_0<0.5$, so a second harmonic amplitude $I_{c2}$ can be half as large to drive the loop hysteretic. For a generic two-component CPR, the non-hysteretic regime is obtained for $\beta_{L2} < 16/((I_{c1}/I_{c2})^2+32)$ for $I_{c1}/I_{c2} \leq 8$ and $\beta_{L2} < 16/((I_{c1}/I_{c2})-2)$ for $I_{c1}/I_{c2} \geq 8$. (see supplemental materials for derivation).

\begin{figure}[ht]
  \includegraphics[width=8cm]{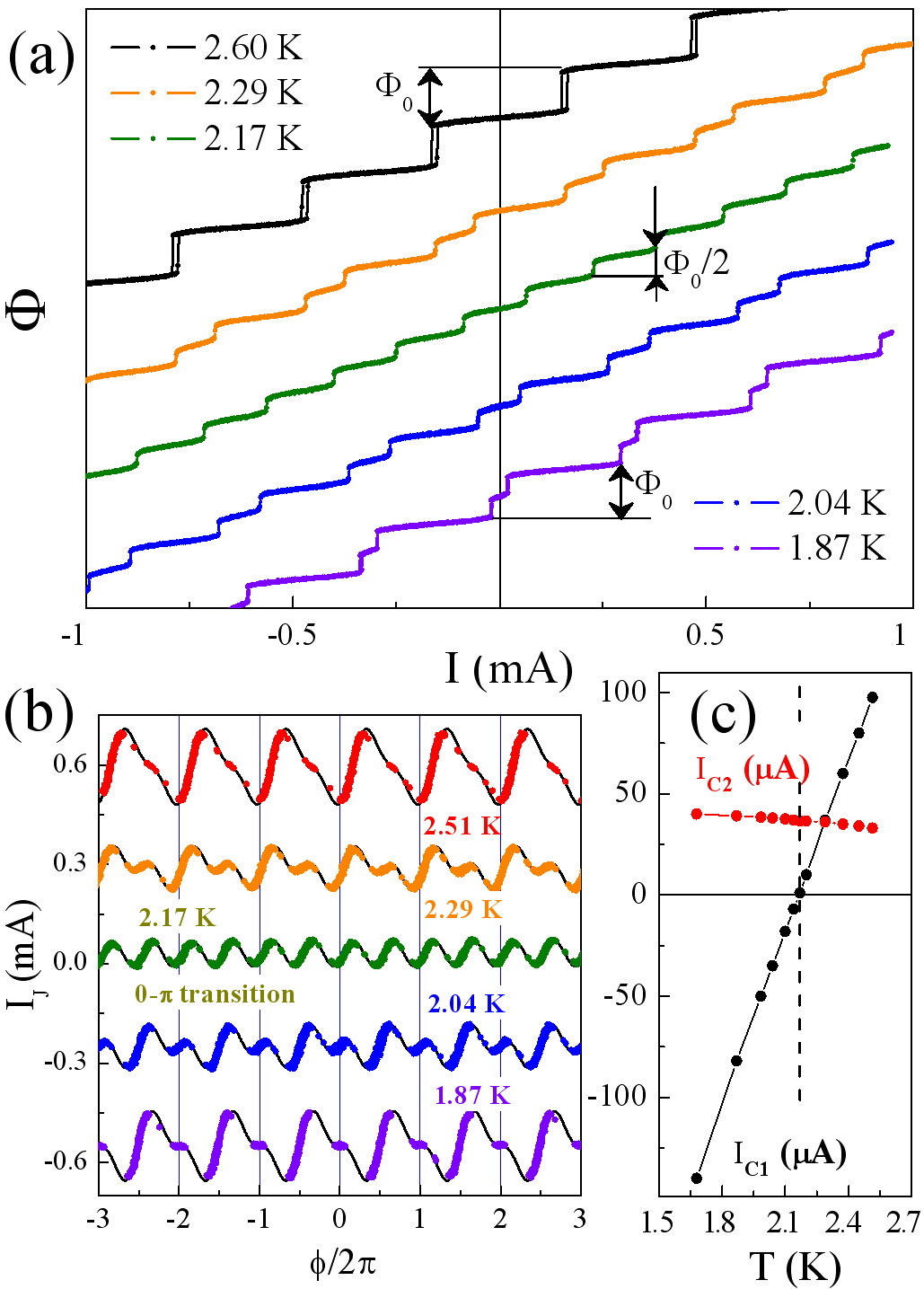}
  \caption{
(a) Readout SQUID signal $\Phi$ as a function of $I$ for a range of temperatures indicated in the legend. $I$ is swept from zero to positive, then to negative, then back to zero (bipolar retrace). 
(b) data in panel (a) without the linear slope due to $L$. Black lines are fits to a two-component CPR. Curves are offset vertically. Horizontal axis scale is based on the periodicity of raw data.
(c) $I_{c1}$ and $I_{c2}$ extracted from fits such as those in panel (a) for an extended set of temperatures. Temperature $T_\pi$ indicated by a vertical dashed line. Solid lines are guides to the eye.
 }
 \label{fig2}
\end{figure}

Fig. 2(a)
shows the readout SQUID signal $\Phi(I)$ for a range of temperatures that includes the 0-$\pi$ transition temperature $T_\pi \approx$ 2.15~K. At T=2.60~K, far above $T_\pi$, a sequence of equidistant steps is observed. This is typical for a weakly hysteretic superconducting loop: near each step, the magnetic flux in the loop abruptly changes by a value close to $\Phi_0$, and the phase across the junction changes by a value close to $2\pi$ (see supplemental materials for a wider temperature range showing strong hysteresis). The overall slope of $\Phi(I)$ corresponds to $L=6.6$~pH and is independent of temperature. At T = 2.29 K the pattern acquires a double-step character, with steps half the height, i.e. close to 0.5$\Phi_0$, occurring at uneven intervals in  $I$. The half-steps become even in a narrow temperature range around $T_\pi$, as shown for T=2.17 K, resulting in half-periodic modulation of $\Phi(I)$. At this temperature the CPR is purely $\mathrm{sin}(2\phi)$.
The characteristics also become less abrupt and more rounded indicating that the loop is approaching the non-hysteretic regime. As the temperature is reduced below $T_\pi$ to 2.04 K and further to 1.87 K, the steps are once again uneven indicating a growing first harmonic. The overall characteristic is shifted by half a period with respect to high temperature curves, this is because the SFS junction has transitioned into the $\pi$-state \cite{Frolov-CPR}.

The current-phase relation is obtained by subtracting a linear contribution due to current in L. 
In Fig. 2(b) we present a series of CPR extracted 
within the temperature range \mbox{$1.7~K~<T<~2.6~K$} where the condition $\beta_{L1}<1$ is fulfilled and half-periodicity is observable. The experimental points in Fig. 2(b) reveal the CPR for a partial range of $\phi$ due to unexpectedly high second harmonic amplitude with $\beta_{L2}=0.7$ leading to weakly hysteretic $\Phi(I)$ dependences even at T = T$_\pi$ (see supplemental materials for details).
Both amplitudes $I_{c1}$ and $I_{c2}$ can be extracted by fitting the experimental $\Phi(I)$ curves assuming a two component CPR in the form $I_J(\phi) = I_{c1}sin(\phi)+I_{c2}sin(2\phi)$, where $I_J$ is the supercurrent through the junction (Fig. 2(c)). We see that the first harmonic crosses zero near $T=T_\pi$, where the CPR becomes $\pi$-periodic. The second harmonic is weakly changing over the entire temperature range and has a positive sign. 
The sign of I$_{c1}$ is fixed to be positive at higher temperatures for SFS junctions with this barrier thickness based on previous studies (Refs.\onlinecite{Oboznov-PRL,bolginovJLTP18}, see also Fig. 5).

Half-periodic CPR extracted from a single-junction loop provides evidence of the dominant second harmonic that is immune to alternative explanations. To further confirm this observation and check it against other common measurements, we cut inductors $L$ and $L_{readout}$ with a focused ion beam (Fig. 1(c)) and perform voltage measurements across the same SFS junction. The same readout SQUID is used, but now in the voltmeter configuration in which $L_{readout}$ and a small standard resistor (20-50 m$\Omega$) are shunting the SFS junction. 

We find a new kind of evidence of the second harmonic CPR in Josephson diffraction, by measuring the critical current as a function of flux threading the junction itself (Figure 3(a)). The diffraction patterns develop a second modulation near $T_\pi$ that is half-periodic in the applied magnetic flux. This striking effect is a confirmation of the presence of a large second Josephson harmonic: indeed in a purely sin(2$\phi$) junction the period of the diffraction pattern should be half the normal period. This is most clearly seen for the applied flux in the range between $-\Phi_0$ and +$\Phi_0$ at T=2.25-2.35 K. To confirm that such diffraction patterns can originate from a two-component CPR, we perform self-consistent simulations of diffraction patterns for a uniform junction, taking as inputs the amplitudes of I$_{c1}$ and I$_{c2}$ from Fig. 2(c) and allowing for a small shift in $T_\pi$ presumably due to the different methods of temperature measurement (see supplemental materials). The simulated curves closely reproduce the experiment (Fig. 3(b)).
The flux axis is calibrated at temperatures T=1.27 K and T = 2.8 K away from $T\pi$, where the diffraction patterns closely follow the Fraunhofer dependence typical for homogeneous Josephson junctions and the CPR is dominated by the first harmonic. All throughout the temperature range of the 0-$\pi$ transition the diffraction patterns exhibit a large peak in the center, at zero applied magnetic flux, thereby confirming that the junctions do not contain significant non-uniformities which would result in a zero-field minimum due to the coexistence of 0- and $\pi$- regions within the junction \cite{Frolov-Shapiro,sickingerPRL12,kemmlerprb10,pfeifferprb08}.

\begin{figure}[ht]
  \includegraphics[width=8.5cm]{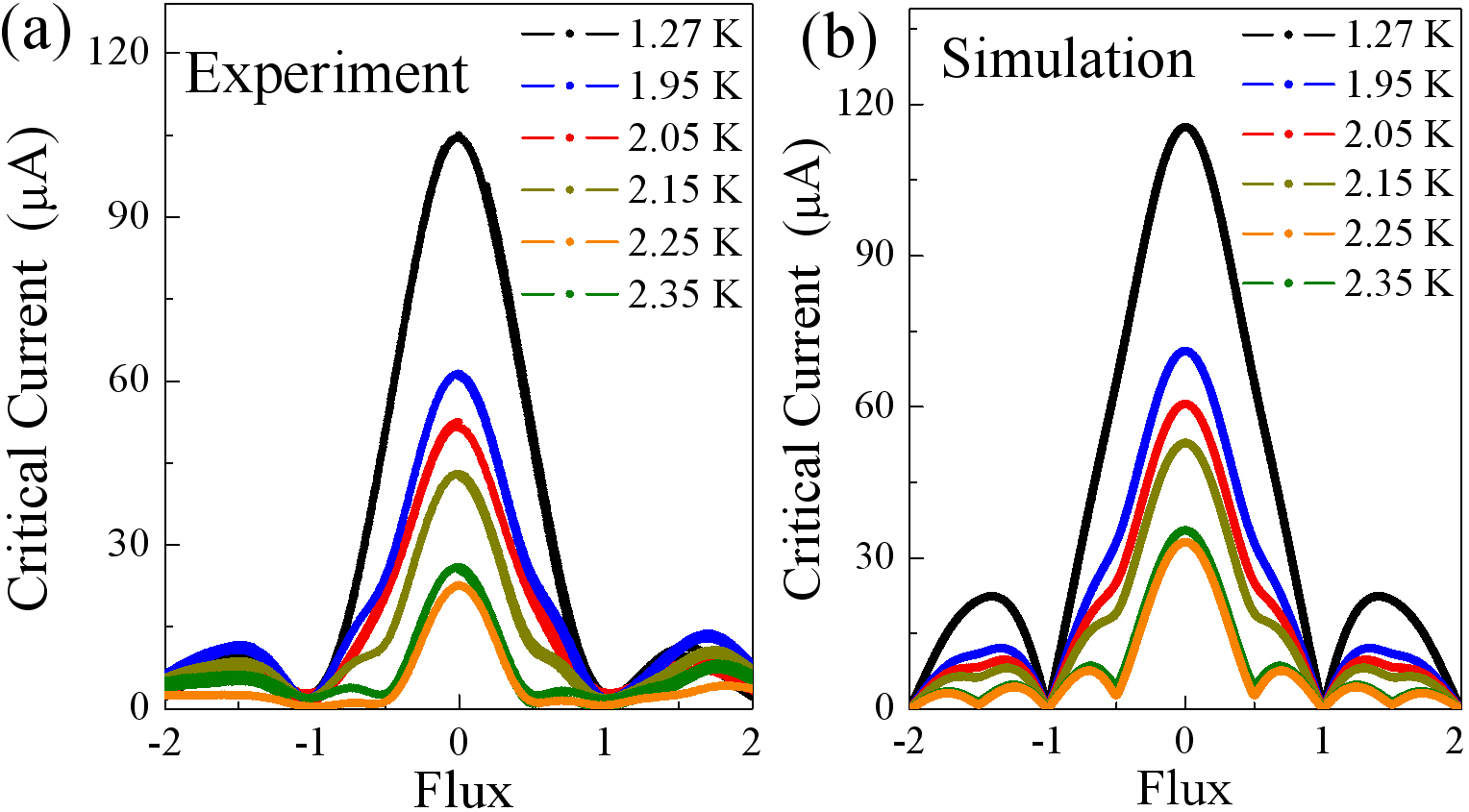}
  \caption{(a) Experimental diffraction patterns for junction studied in Fig. 2 over a range of temperatures. (b) Simulated diffraction patterns using CPR with $I_{c1}$ and $I_{c2}$ from Fig. 2(c). Colors correspond to temperatures in the legend of panel (a).
 }
 \label{fig3}
\end{figure}

\begin{figure}[ht]
  \includegraphics[width=\columnwidth]{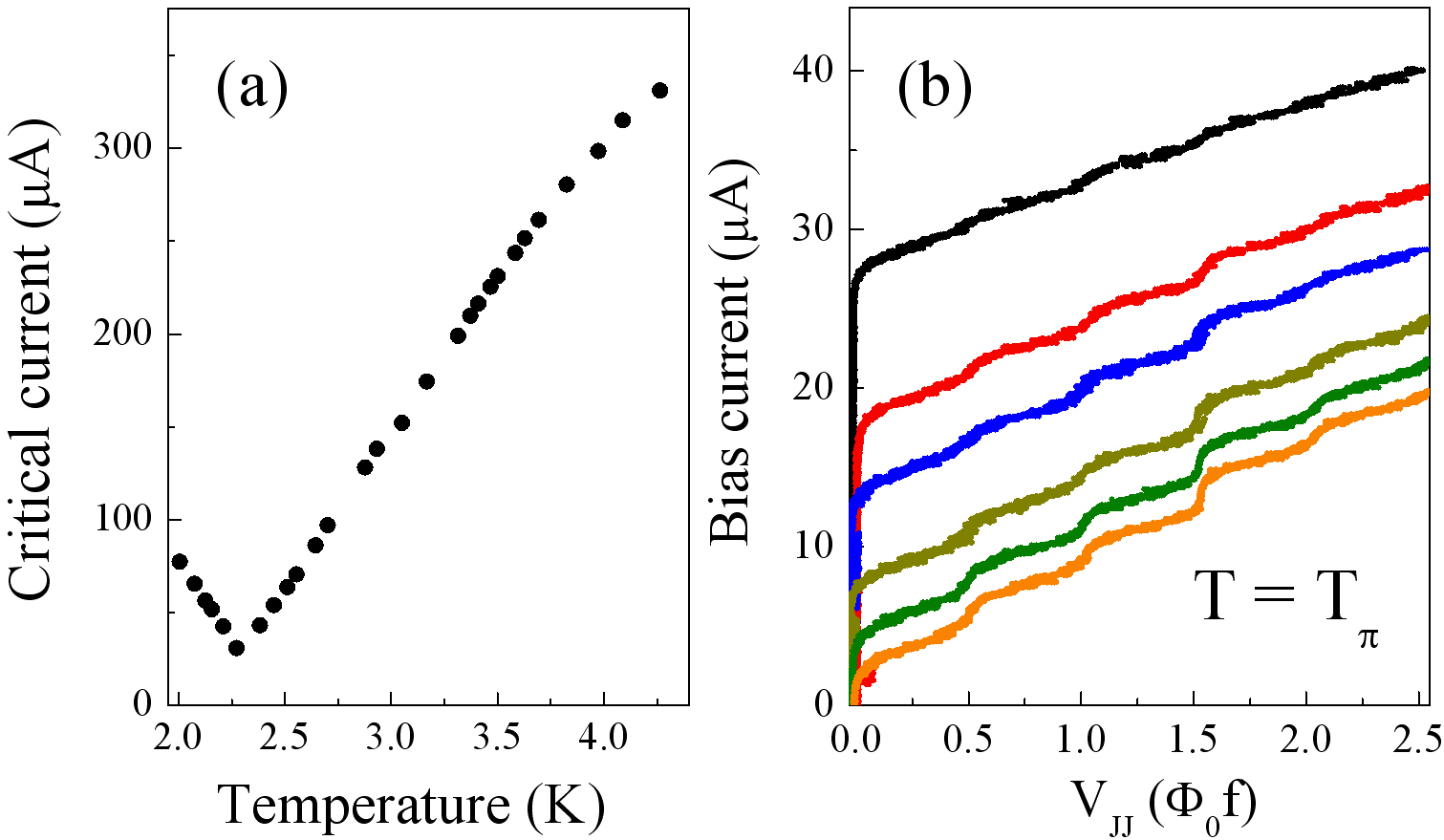}
  \caption{(a) Critical current vs. temperature for junction studied in Figs. 2 and 3. (b) Example half-integer Shapiro steps for a range of applied rf power (frequency 1.68 MHz) with black trace at lowest power and orange trace at highest applied power.
 }
 \label{fig4}
\end{figure}

Phase-sensitive measurements (Figs. 2 and 3) are in agreement with transport measurements (Fig. \ref{fig4}). The temperature dependence of the total critical current $I_c$ for the same junction is plotted in Fig. 4(a). The data show a steady decrease of $I_c$ as the temperature is lowered down to $T=T_\pi$. 
Below $T = T_\pi$, the critical current increases \cite{Ryazanov-junction,Aprili-junction}.
At $T=T_\pi$ $I_c$ not reach zero, saturating at $I_c \approx$~ 30~$\mu A$. 
This value is consistent with $I_{c2}$ extracted from CPR measurements.

Shapiro step measurements are also commonly used to identify non-sinusoidal CPRs \cite{Sellier-Sin2phi, wiedenmann20164pi}. In this measurement, the junction is excited with an ac signal at frequency $f$. Shapiro steps appear at voltages $V_{jj}$ equal to integer values nf$\Phi_0$ for the first Josephson harmonic and half-integer values (nf$\Phi_0$/2) for the second harmonic. Fig. 4(b) shows examples of the junction current-voltage characteristics with ac excitation applied near $T_\pi$. There are steps at both integer and half-integer multiples of f$\Phi_0$.  Data from additional single junction samples are presented in supplemental materials, confirming the findings.

\begin{figure}[ht]
  \includegraphics[width=8.5cm]{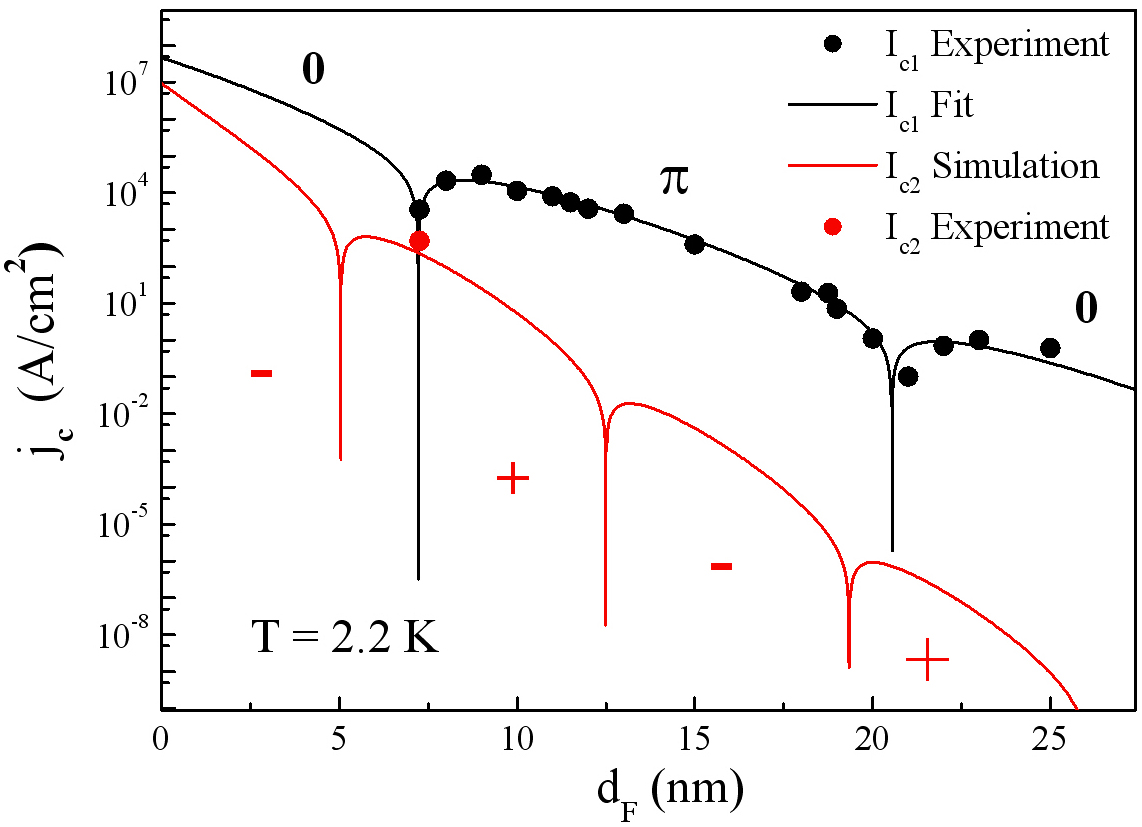}
  \caption{Experimental barrier thickness dependence of critical current density $j_c(d_F)$ (black circles) at T = 2.2 K for junctions fabricated using the trilayer process. The data were obtained during experiment described in Ref.[\onlinecite{bolginovJLTP18}]. The second harmonic current density from Fig. 2 is shown as a red circle. Black and red lines  are fits for $j_{c1}$ and $j_{c2}$ based on Ref.[\onlinecite{BuzdinPRB05}]. The key fit parameters are the critical current density at zero temperature $j_0=5\times10^7A/cm^2$, the critical current decay length $\xi_{F1}=1.3$ nm and the oscillation length $\xi_{F2}$ = 4.3 nm.}
 \label{fig5}
\end{figure}

We comment that conclusions about the second harmonic cannot be made based alone on the 
transport 
measurements
presented 
in Fig.\ref{fig4}. The non-vanishing $I_c$ accompanied by half-integer Shapiro steps were interpreted in the past as evidence of a CPR dominated by sin(2$\phi$) near T=T$_\pi$ \cite{Sellier-Sin2phi}. However, an alternative explanation for non-vanishing critical current is due to
step-like 
barrier inhomogeneities \cite{Frolov-Shapiro}. In this case, the junction can break into segments that have already transitioned into the $\pi$-state and segments that remain in the 0-state. To satisfy phase continuity, supercurrents circulate in this mixed 0-$\pi$ regime around the F-layer causing a non-vanishing $I_c$. Half-integer Shapiro steps then appear due to phase-locking of these spontaneous supercurrents to the ac excitation\cite{Vanneste, Frolov-Shapiro}.
Therefore phase-insensitive measurements (Fig. 4) have to be supplemented by phase-sensitive measurements of the type presented in Fig. 2 or Fig. 3. 


The last possibility to discuss is related to fine-scale barrier inhomogeneities. In this case a junction can demonstrate I$_c$(H) patterns similar to Fig.3(b) with maximum critical current at zero magnetic flux, but the sign of second harmonic is predicted to be negative \cite{BuzdinKoshelev:2003,GoldobinPRB07}. This origin is not relevant to our samples since our rf-squid based experiments (see Fig. 2) have revealed a positive sign of the second harmonic. The positive sign is in agreement with a prediction from a microscopic theory for diffusive SFS-junctions with uniform ferromagnetic barrier \cite{BuzdinPRB05}

In order to understand the large magnitude of the second order term we fitted the thickness dependence of the critical current density for trilayer SFS junctions to the microscopic theory \cite{BuzdinPRB05}. In Fig. 5, the experimental data show sharp dips at barrier thicknesses $d_F$~=~7.5~nm and 21.5~nm, which are the thicknesses of the first and second 0-$\pi$ transitions. We first fitted these data assuming a purely first harmonic (black line). This allowed us to obtain the key fitting parameters. These parameters have been substituted into the analytical formula for the second harmonic \cite{BuzdinPRB05}. The second harmonic is generally orders of magnitude smaller (red line), however it can dominate at the first 0-$\pi$ transition.
The theoretical amplitude of the second order term at $d_F$ = 7.5 nm is estimated to be 230 A/cm$^2$ which is close to the experimental values (see red dot in Fig. 5). 
Note that 
the theoretical result [\onlinecite{BuzdinPRB05}]  was obtained 
near the critical temperature  of superconducting electrodes. Therefore it cannot provide the exact quantitative coincidence with the experimental value at T~=~2.2 K far from the niobium critical temperature.
Previous CPR measurement on a similar SFS junction \cite{Frolov-CPR} was performed at much larger $d_F$ (22 nm) near the second 0-$\pi$ transition point and showed a purely first order CPR. The second harmonic term expected from the theory in Ref.[\onlinecite{BuzdinPRB05}] for this thickness is about $10^{-6} A/cm^2$ which is too small to be measured.


In conclusion, we have demonstrated a Josephson junction with a 
$\pi$-periodical 
current-phase relation. The regime occurs at the 0-$\pi$ transition of a superconductor-ferromagnet-superconductor junction. Alternative explanations are ruled out by comparing results from four independent methods, both phase-sensitive and transport ones, 
that point at a significant $
\mathrm{sin}(2\phi)$ term. 
While the measured second-order term is surprisingly large for a diffusive SFS junction, it is in agreement with a recent theory. 
The findings and methodology presented here can be used to evaluate exotic current-phase relations 
of other important systems, such as ballistic and topological Josephson junctions.

Acknowledgements. Authors thank N.S. Stepakov and V.N. Shilov for assistance in sample fabrication and measurements. 
M.J.A.S and D.J.V.H are supported by NSF DMR-1710437 and NSF DMR-1610114. S.M.F. is supported by NSF PIRE-1743717, NSF DMR-1252962, NSF DMR-1743972, and ONR. V.V.B. and A.N.R. are supported by Russian Foundation for Basic Research grant 17-02-01270. V.V.R. is supported by the Ministry for Education and Science of Russian Federation under contract no. K2-2014-025. Support from the TZ-93 Basic Research Program of the National Research 
University Higher School of Economics is gratefully acknowledged by NP.

\bibliographystyle{apsrev4-1}
\bibliography{sfs.bib}

\clearpage
\setcounter{figure}{0}
\renewcommand{\thefigure}{S\arabic{figure}}
\subsection*{SUPPLEMENTAL MATERIALS}

\begin{figure*}[b]
\centering
  \includegraphics[width=\textwidth]{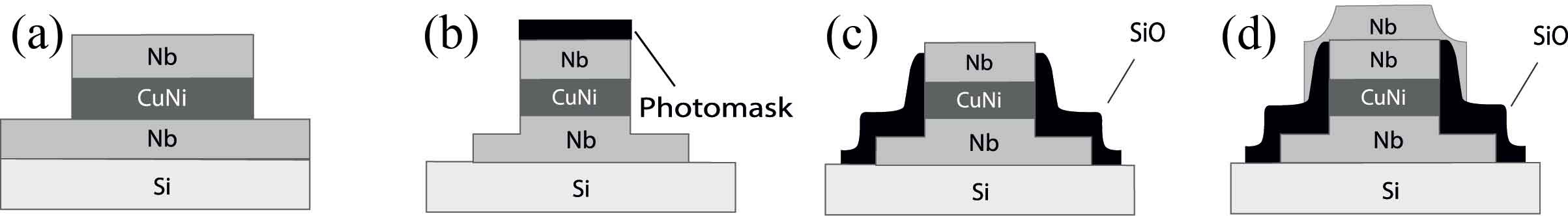}
  \caption{The self-aligned trilayer fabrication process. 
}
 \label{figS1}
\end{figure*}

\textbf{Self-aligned trilayer fabrication process}. The fabrication process consists of the following main steps. 
\begin{enumerate}
\item Deposition of a Nb-CuNi-Nb trilayer on a Si substrate coated with a thin AlO$_x$ layer.
\item  Optical lithography to form the bottom superconducting electrode and subsequent argon etching of two upper layers of the trilayer (Fig. \ref{figS1}(a)).
\item  Optical lithography to form SFS mesa followed by argon etching. At this step the bottom electrode and the superconducting contact pads are also formed. The photoresist mask is retained at the end of this step (Fig. \ref{figS1}(b)).
\item Thermal deposition of SiO followed by a lift-off process (Fig. \ref{figS1}(c)).
\item Thermal deposition of additional SiO isolation and lift-off process using pre-created photoresist mask (not shown in Fig. \ref{figS1}).
\item Wiring layer fabrication by means of optical lithography followed by Nb magnetron sputtering and lift-off process (Fig. \ref{figS1}(d)). The ion cleaning of the substrate is performed before the deposition to ensure good interface transparency to the top niobium layer.
\end{enumerate}

In general our process is identical to that described in Ref.[\onlinecite{bolginovJLTP18}]. Two substantial additions have been made. First, the secondary isolation layer is fabricated between the overlapping bottom electrode and wiring parts within the readout loop (see Fig. S2). The overlap allows excluding stray magnetic flux (for example, from shunting to readout loop) and thus simplifies the subsequent data analysis. The additional isolation helps us avoid shorts. Second, approximately 7 percent of oxygen is added to argon during argon plasma etching at the third stage. This guarantees that etched niobium is immediately oxidized and can't create any superconducting or metallic shorts at the mesa boundary. All other details coincide with that in Ref.[\onlinecite{bolginovJLTP18}]. 

\textbf{Temperature control and measurement}. Experiments reported in the work were done in He-4 pumped cryostats. The temperature was varied in the range 1.27~K~$<T<$~4.2~K and controlled by adjusting the He-pressure in the cryostat using a membrane valve. For the CPR experiments the vapor pressure was used for thermometry, while in transport measurements an on-chip silicon diode thermometer was used. This difference likely accounts for a 0.1 K shift in the recorded $T_\pi$ between Fig. 2 and Fig. 4 of the main text.

\textbf{Derivation of non-hysteresis condition for rf-SQUID in the case of two-component CPR}. 

The current-phase relation can be extracted from experimental data (see Fig. 2) if $\Phi(I)$ is a single-valued function. This imposes the restriction on the loop inductance and the junction critical current. To derive this, we start with the expression for supercurrent in a single-junction loop supplemented with phase balance condition for rf-SQUID \cite{Barone}:

\[
I_{s}=\frac{\Phi }{L}+I_{c1}\sin \phi +I_{c2}\sin 2\phi 
\]
\[
\phi = 2\pi\Phi/\Phi_0
\]

and require the derivative $dI_{s}/d\Phi $ to be positive for every value of $\Phi$. As a result, we obtain the following condition which must be satisfied for every value of $\phi$:

\[
1+G\beta_{L2} \cos \phi +2\beta_{L2} \cos 2\phi \geqslant 0,
\]

where $G=$$I_{c1}/I_{c2}$. We express the singlevaluedness condition in terms of G and $\beta_{L2}$ because experimentally $I_{c2}$ and hence $\beta_{L2}$ have a very weak temperature dependence near the 0-$\pi$ transition. From the above we get:

\begin{eqnarray*}
1+G\beta_{L2} \cos \phi +4\beta_{L2} \cos ^{2}\phi -2\beta_{L2}  &\geqslant &0, \\
\frac{1-2\beta_{L2} }{4\beta_{L2} }-\frac{G^{2}}{64}+\left( \cos \phi +\frac{G}{8}
\right) ^{2} &\geqslant &0.
\end{eqnarray*}

In this form, only one term is $\phi$-dependent. We study two regimes with respect to G. First, let $G\leqslant 8$. Then the minimal value of $\left( \cos \phi +\frac{G}{8}\right)^{2}$ is zero. The singlevaluedness condition reduces to:

\[
\frac{1-2\beta_{L2} }{4\beta_{L2}}-\frac{G^{2}}{64}\geqslant 0
\]

Or:

\[
\beta_{L2} \leqslant \frac{16}{G^{2}+32}.
\]

Second, for $G>8$ we have for every $\phi$:%
\[
\left( \cos \phi +\frac{G}{8}\right) ^{2}\geqslant \frac{\left(
G-8\right) ^{2}}{64}
\]%
Therefore the singlevaluedness condition becomes:%
\begin{eqnarray*}
\frac{1-2\beta_{L2} }{4\beta_{L2} }-\frac{G^{2}}{64}+\frac{G^{2}}{64}-\frac{G}{4}+1
&\geqslant&0 \\
1+2\beta_{L2} -G\beta_{L2}  &\geqslant&0 \\
\beta_{L2}  &\leqslant&\frac{1}{G-2}
\end{eqnarray*}

at $G\geqslant 8$.

The dependence of the critical value of $\beta_{L2} $ on G is plotted at Fig. \ref{figS7}. One can see that the maximum value of $\beta _{L2}^{crit}$ is 0.5 which is achieved for G = 0. In our experiment $\beta_{L2}$ at $T = T_{\pi}$ was estimated to be 0.7 and therefore the experimental $\Phi(I)$ curve is never single-valued. One should note that the single-valuedness condition for rf-SQUID with a two-component CPR junction is stricter than for a single-component CPR junction. As G grows due to increasing first harmonic component, the condition on $\beta_{L2}$ further tightens. We also note that in experiment in Fig. 2 of the main text, the hysteresis of CPR curves, caused by the second harmonic component, is largely suppressed near $T_\pi$. This is due to thermal fluctuations and/or flux noise that induce jumps between metastable states.

\begin{figure*}
  \includegraphics[width=12cm]{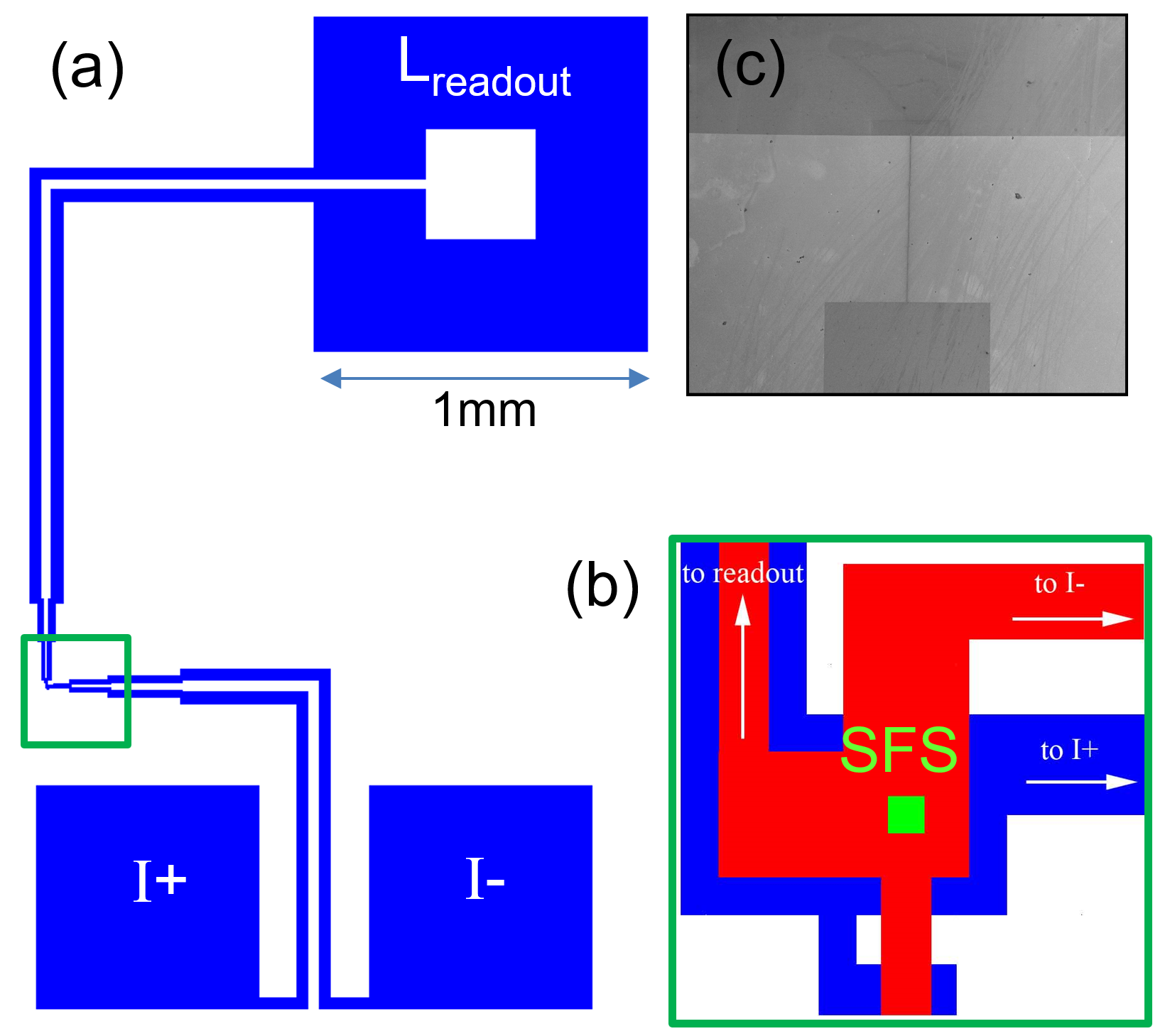}
  \caption{(a) layout of the full device with Nb films in blue. Shown are the washer for gluing a pickup coil of a commercial SQUID sensor ($L_{readout}$) as well as current injection pads (I+ and I-). The green square marks the location of the SFS junction expanded in panel (b), in which the top and bottom Nb layers are blue and red respectively, green square is the location of the SFS trilayer. Red and blue Nb electrodes form the inductor $L$ shown below the SFS junction, with dimensions 2x4 $\mu m^2$ (inner)  and 9x8 $\mu m^2$ (outer). (c) Optical image of the washer with a focused ion beam cut, to complement Fig. 1(c) in the main text. 
  }
 \label{figS2}
\end{figure*}

\begin{figure*}
  \includegraphics[width=11cm]{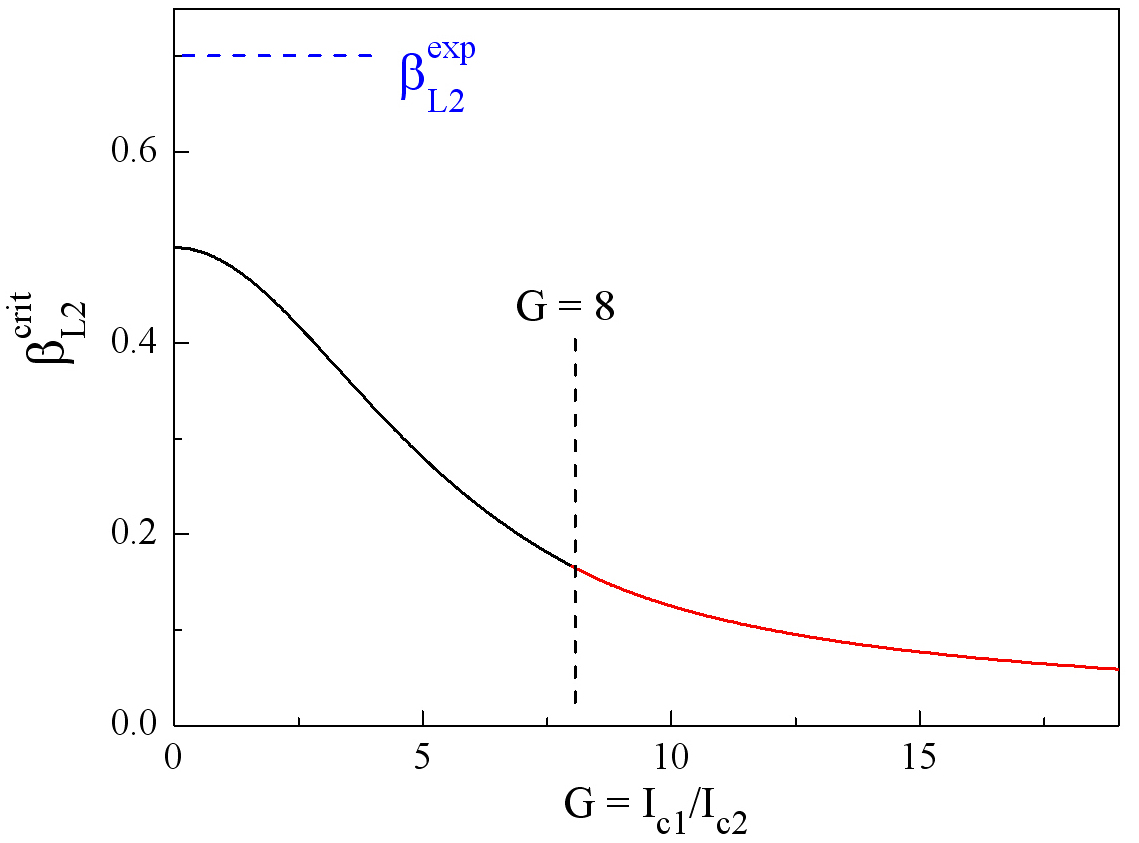}
  \caption{Analytical plot of $\beta^{crit}_{L2}$, the critical value of $\beta_{L2}$ above which the single junction loop (rf SQUID) is hysteretic for different values of parameter $G$. A horizontal dashed line marks the value of $\beta^{exp}_{L2}$ experimentally measured for the junction studied in the main text based on the magnitude of $I_{c2}$ and the loop inductance $L$. The vertical dashed line marks transition between two regimes used in the derivation of this dependence.
 }
 \label{figS7}
\end{figure*}

\begin{figure*}
  \includegraphics[width=12cm]{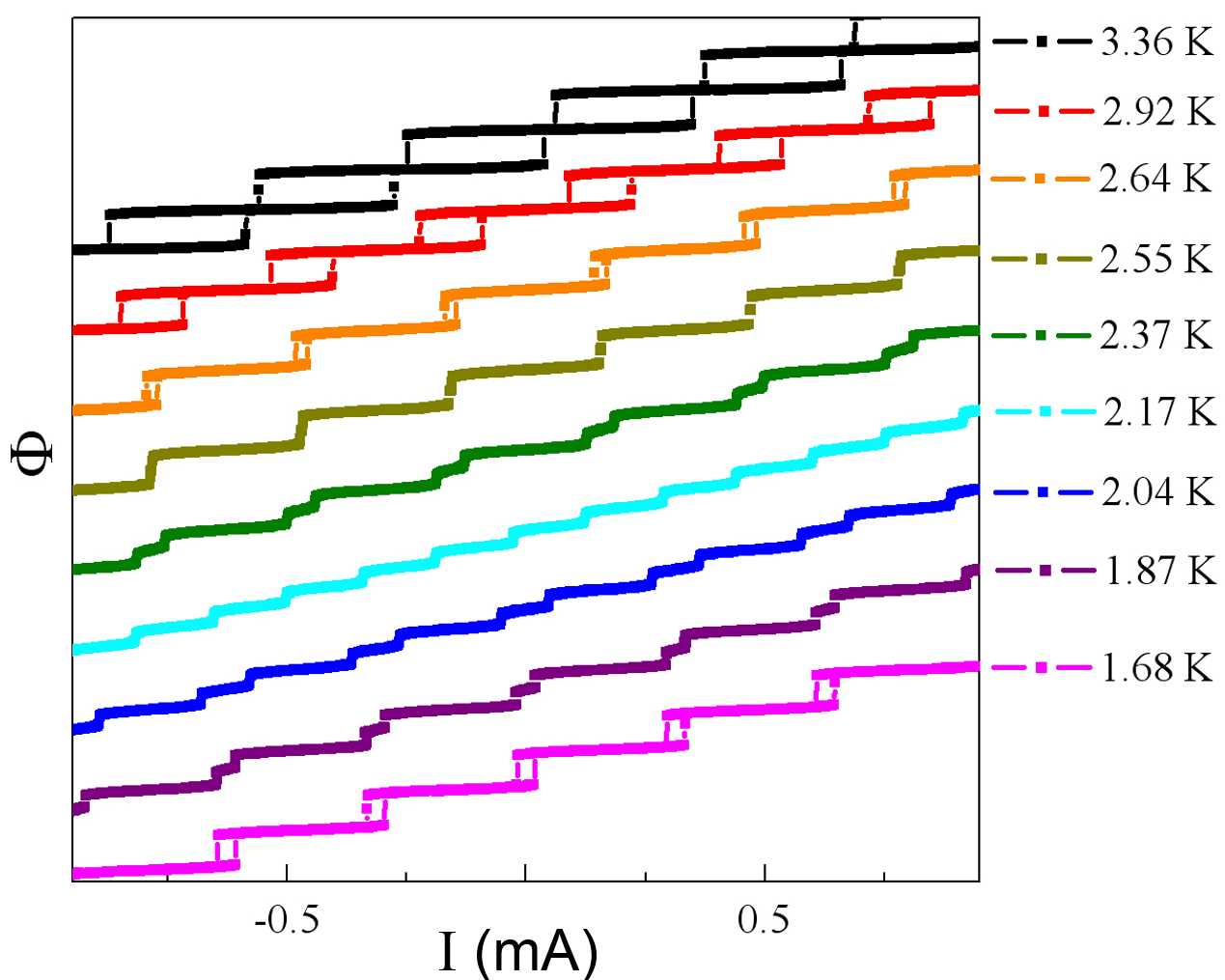}
  \caption{Raw CPR experimental data in a wider range of temperatures, complementary to data in Fig.\ref{fig2} in the main text. The curves are shifted for better view.
 }
 \label{figS3}
\end{figure*}

\begin{figure*}
  \includegraphics[width=12cm]{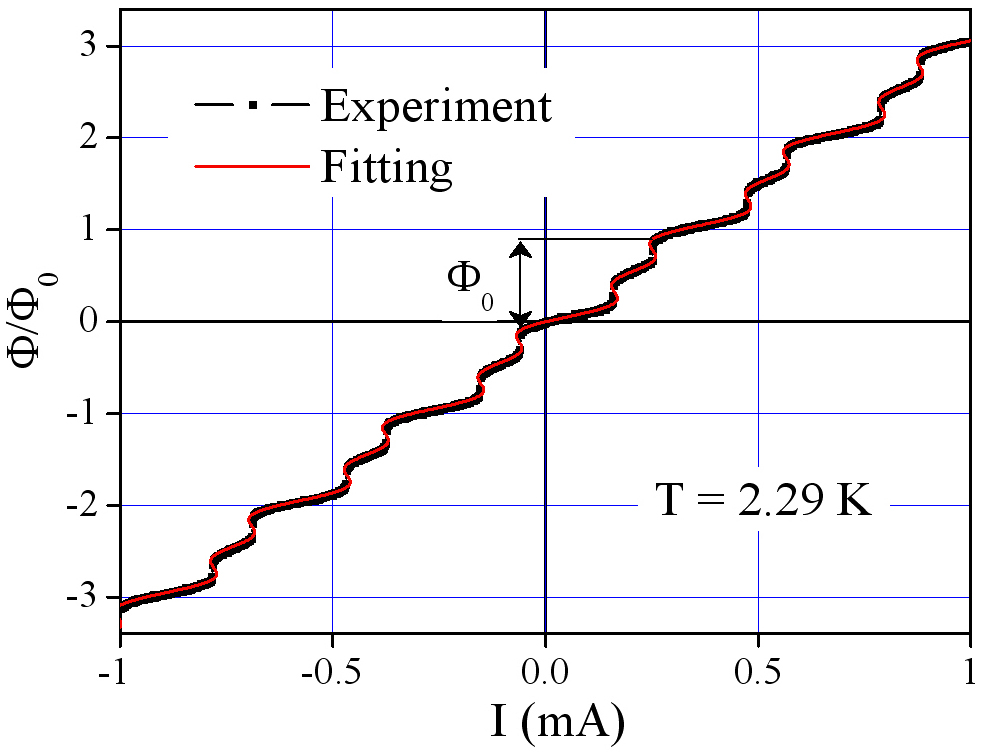}
  \caption{Example of raw CPR data (black) with a two-component CPR fit (red).
 }
 \label{figS4}
\end{figure*}

\begin{figure*}
  \includegraphics[width=14cm]{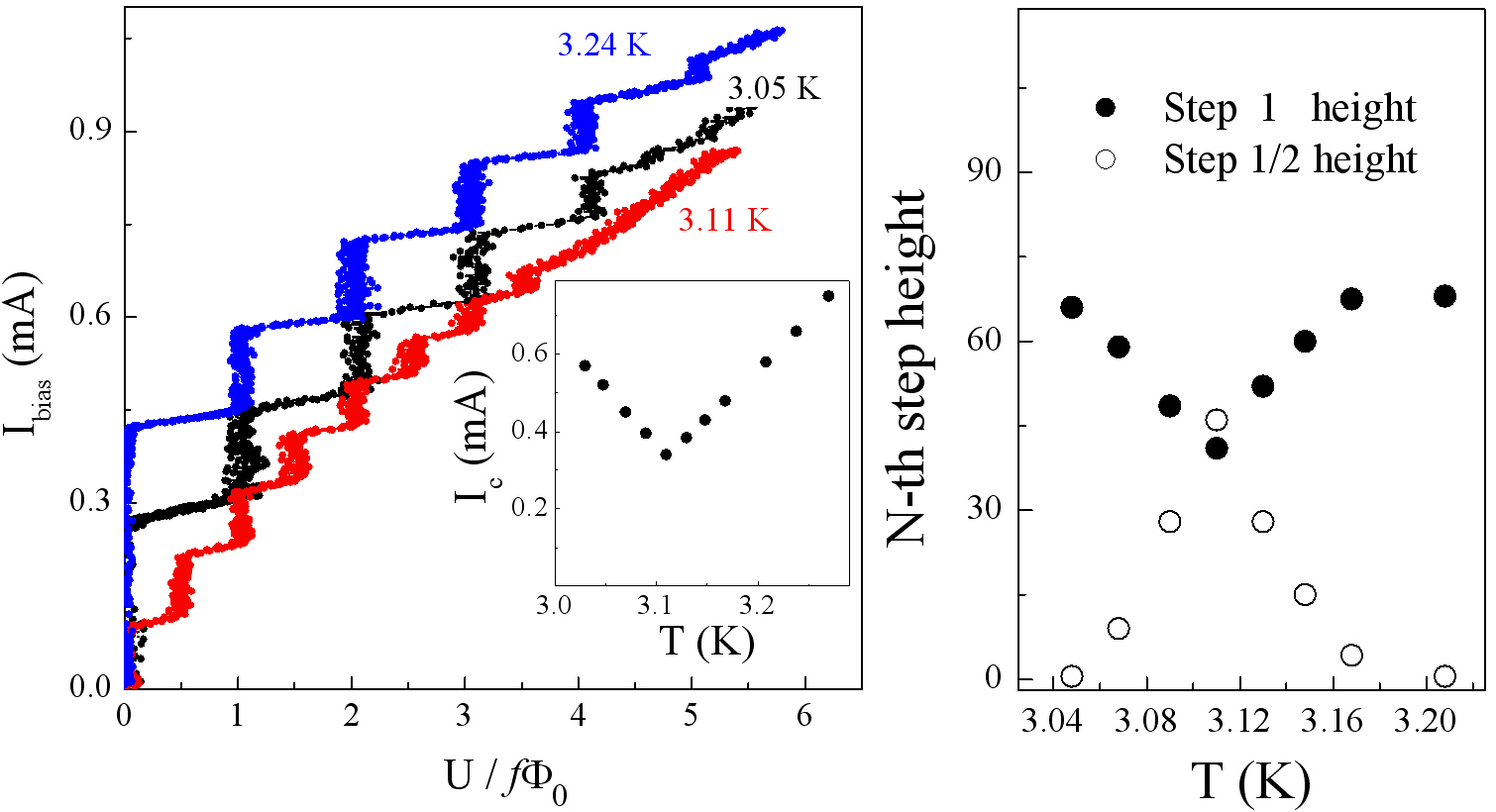}
  \caption{Transport data from a $10\times 10\ \mu \mathrm{m}^2$ SFS junction. Prior to CPR investigations, up to 10 single junctions were studied showing effects consistent with those reported in Figs. 3 and 4 of the main text. (a) Half-integer Shapiro steps appearing at $T = T_{\pi}$ (3.11~K for given junction). AC-current frequency is 7~MHz. The inset shows the temperature dependence of the critical current from this junction. Higher critical current due to larger junction area resulted in shaper integer and half-integer steps. (b) Temperature dependence of the amplitudes of the first integer and half-integer Shapiro steps for the same power applied. The half-integer steps appear in a narrow temperature range of order 0.1 K near $T_\pi$. 
 }
 \label{figS5}
\end{figure*}

\begin{figure*}
  \includegraphics[width=14cm]{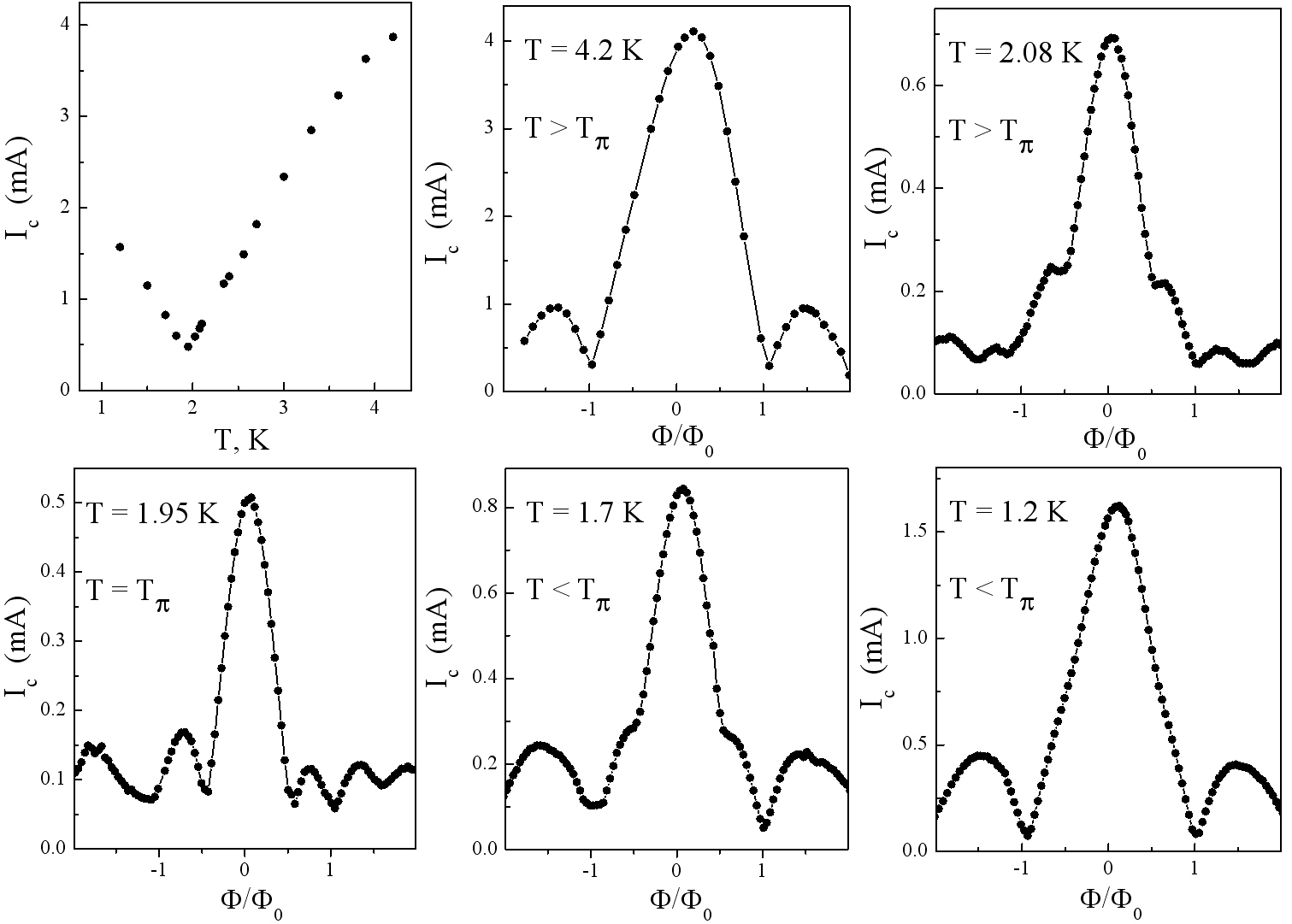}
  \caption{$I_c(T)$ and diffraction patterns at four temperatures obtained from one more $10\times 10\ \mu \mathrm{m}^2$ SFS junction. The dependence is close to a Fraunhofer diffraction pattern far away from $T_\pi=1.95$~K while local peculiarities in critical current occurs at half-integer flux near $T_\pi$. Small asymmetry near $ T_{\pi}$ is due to weak magnetic inhomogeneity of the CuNi barrier. 
 }
 \label{figS6}
\end{figure*}

\end{document}